\documentclass[%
 reprint,
superscriptaddress,
amsmath,amssymb,
aps,
floatfix,
]{revtex4-2}

\usepackage[english]{babel}
\usepackage{physics}

\usepackage{amsmath,amssymb,amsthm}
\usepackage{tikz}

\usepackage[flushmargin]{footmisc}
\renewcommand{\footnoterule}{%
  \kern-3pt
  \hrule width \textwidth height 0.4pt
  \kern 2.6pt
}
\usepackage{physics}
\usepackage{graphicx}
\usepackage[colorlinks=true, allcolors=blue]{hyperref}

\definecolor{blau}{RGB}{50,50,150}

\begin{document}
\preprint{APS/123-QED}

\title{A quantum state transfer protocol with Ising Hamiltonians}

\author{Oscar Michel}
\email{oscar.michel@qilimanjaro.tech}
\affiliation{Qilimanjaro Quantum Tech., Carrer de Veneçuela, 74, Sant Martí, 08019 Barcelona, Spain}
\affiliation{Departament de Física Quàntica i Astrofísica (FQA), Universitat de Barcelona (UB), Carrer de Martí i Franqués, 1, 08028 Barcelona, Spain}

\author{Matthias Werner}
\email{matthias.werner@qilimanjaro.tech}
\affiliation{Qilimanjaro Quantum Tech., Carrer de Veneçuela, 74, Sant Martí, 08019 Barcelona, Spain}
\affiliation{Departament de Física Quàntica i Astrofísica (FQA), Universitat de Barcelona (UB), Carrer de Martí i Franqués, 1, 08028 Barcelona, Spain}
\affiliation{Institut de Ci\`encies del Cosmos, Universitat de Barcelona, ICCUB, Carrer de Martí i Franqu\`es, 1, 08028 Barcelona, Spain}

\author{Arnau Riera}
\email{arnau.riera@qilimanjaro.tech}
\affiliation{Qilimanjaro Quantum Tech., Carrer de Veneçuela, 74, Sant Martí, 08019 Barcelona, Spain}

\begin{abstract}
    Quantum state transfer is a fundamental requirement for scalable quantum computation, where fast and reliable communication between distant subsystems is essential. In this work, we present a protocol for quantum state transfer in linear Ising chains. Starting from a perfect state transfer scheme via a Heisenberg Hamiltonian with inhomogeneous couplings, we adapt it for architectures implementing the transverse-field Ising model by encoding the information in domain walls. The resulting linear Ising chain makes quantum transport experiments accessible to many platforms for analog quantum simulation. We test the protocol for 1-, 2-, and 3- spin states, obtaining high transfer fidelities of up to 0.99 and study the accuracy dependence on the domain wall approximation. These results are the first step in paving the way for an experimental implementation of the protocol.

\end{abstract}

\maketitle

\section{Introduction}
\label{sec:introduction}

Analog quantum computing has in recent years become an area of great interest for researchers and industry, for its many applications that include simulation of condensed matter systems \cite{king_coherent_2022, kyriienko_floquet_2018}, quantum machine learning \cite{kornjaca_large-scale_2024}, and optimisation problems \cite{farhi_quantum_2000, das_quantum_2008, leng_differentiable_2022}. Additionally, the technology underlying analog quantum computers is at a stage where devices can be built with a large quantity of precise and coherent qubits, enabling the realisation of the aforementioned applications, either at present or in the near-term.

One challenging aspect of running quantum algorithms in analog devices is the transfer of information within the Quantum Processing Unit (QPU), since the copying of information is barred by the no-clone theorem \cite{wootters_single_1982}, and the SWAP operation typically performed in digital devices is not necessarily available on all quantum platforms. It becomes a problem of interest to find quantum state transfer protocols for these cases, both as a working piece of a larger quantum information processing system as well as a system to be studied in its own right.

The core objective of a quantum state transfer (QST)–or quantum transport–protocol is to transfer a quantum state across a certain distance with as high fidelity as possible. Simulating quantum transport in analog devices can allow the study of long-range communication and multipartite state transfer in chains of qubits and is applicable not only to intra-processor information transfer \cite{zhou_high_2014}, but also to problems such as long-range quantum communications \cite{khabiboulline_optical_2019} and quantum metrology \cite{muralidharan_optimal_2016}. Additionally, a simple state transfer protocol can be used as a benchmark test for small devices to measure their accuracy in quantum simulation, the level of control over individual qubits, and the errors introduced by hardware defects.

There are many proposals to create protocols for efficient and realizable quantum transport \cite{christandl_perfect_2004, wojcik_unmodulated_2005, burgarth_conclusive_2005, gualdi_entanglement_2009}. However, implementing them on hardware poses both technical and theoretical difficulties. It is therefore very relevant to determine which transport schemes work best in different devices by studying their limitations and simulating them in analog hardware. This is precisely the goal of this paper, where we explore strategies to implement quantum transport in devices that natively realise Ising Hamiltonians. Some examples of such systems are coupled superconducting flux qubits \cite{DWaveTechReport, King_2025} and Rydberg atom arrays \cite{Henriet_2020, Wurtz_2023}.
 
The main limitation of previous protocols is given by the hardware they are applied on, which in many cases restricts us to modeling an Ising-like system with a transverse field. More exotic interactions, such as XX and YY terms are for many platforms still in an early stage of development \cite{ozfidan_demonstration_2020, hita-perez_ultrastrong_2022}. This is in principle a hard restriction, since most transfer protocols work with Heisenberg Hamiltonians and involve interactions not yet available e.g. in many superconducting devices. However, the choice of information encoding can expand the range of systems that we can recreate. In particular, we will perform simulations of quantum transport algorithms using a domain wall encoding \cite{chancellor_domain_2019}, which will allow us to overcome some of the hardware restrictions and simulate interactions beyond the ZZ term in one-dimensional systems. With this strategy we aim to achieve near-perfect single-qubit, and multi-qubit state transfer. 
 
The content of this paper will be structured as follows: Section \ref{sec:standard_transfer} introduces the problem of quantum state transfer in more detail, as well as the original strategy by Christiandl et al. \cite{christandl_perfect_2004} on which we have based our protocol. Section \ref{sec:domain_walls} contains the main characteristics of domain wall encoding and introduces the Hamiltonian that will be used for our version of the transfer protocol. Section \ref{sec:DW_protocol} describes the adaptation we have made to implement the protocol with domain walls and gives examples of state transfer with this strategy. Furthermore, we analyze the precision of the domain wall approximation, before ending with the conclusions and possible extensions in Section \ref{sec:conclusions}.

\section{Quantum state transfer}
\label{sec:standard_transfer}

In this paper we will simulate quantum state transfer along one-dimensional chains of spin-1/2 particles. Here, we use the words spin and qubit interchangeably with the identification $\ket{0} \equiv \ket{\downarrow}$ and $\ket{1} \equiv \ket{\uparrow}$. The basic rules of QST are that the sender (Alice) and the receiver (Bob) have full control of their local qubits, or more generally, their local register of qubits. Between them they have a quantum wire that initially is in a pre-defined state. Alice prepares the state they wish to send in their register and the system is left to evolve under a given Hamiltonian, which must not depend on the state that Alice wants to send. In principle, the global system, i.e. registers and the wire, are allowed to be changed, the changes, however, must not depend on the quantum data that is transported. An exception, obviously, is Alice when loading the quantum data to their register.

\begin{figure}
    \centering
    \includegraphics[width=0.475\textwidth]{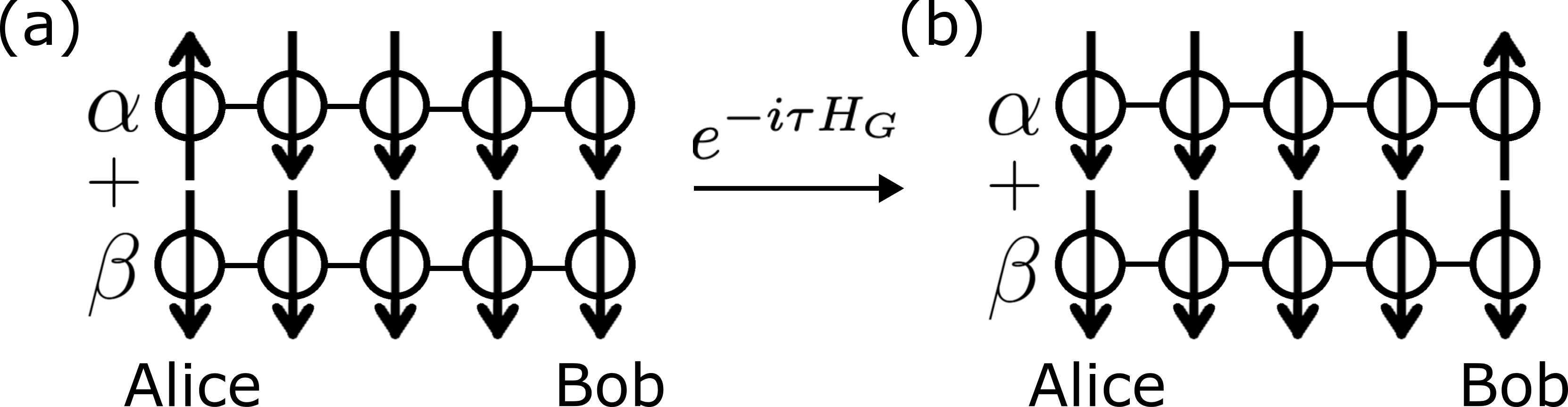} 
    \caption{Schematic representation of the initial state (a) that evolves under $H_G$ Eq. \eqref{eq:hamiltonian_standard} into the final state (b) of the transport process of $\ket{\psi} = \alpha\ket{1} + \beta\ket{0}$ along an example chain of 5 spins.}
    \label{fig:diagram_standard}
\end{figure}

The basics of the transfer process are the following. We start with a chain of $N$ spins, all in the state $\ket{0}$, and Alice prepares a spin in their end in the state that they wish to transfer. This is generally a superposition 

\begin{equation}
    \ket{\psi (t=0) }_1 = \alpha\ket{1} + \beta\ket{0} \ ,
\end{equation}
where $\alpha, \beta \in \mathbb{C}$, $|\alpha|^2 + |\beta|^2 = 1$. In a chain of spins arranged from left to right, we will encode the state in the leftmost spin, and will refer to it as the first spin in the chain. The transfer process consists of applying a given Hamiltonian to the chain and evolving it over time. We consider that the state has been successfully transferred, when at a certain time $\tau$ all the spins in the chain are in the state $\ket{0}$, except for Bob's spin, the last spin in the chain, which now is in the same state as the Alice's spin at time $t=0$, i.e.

\begin{equation}
    \ket{\psi (t = \tau)}_N = \ket{\psi (t=0) }_1 = \alpha\ket{1} + \beta\ket{0} \ .
\end{equation}
A visual representation of this process can be seen in Figure \ref{fig:diagram_standard}.

The main complication of this process is the choice of Hamiltonian, since we require very specific conditions for the state transfer to be successful. Over the years, several candidates have emerged, each with its own advantages and drawbacks. We can summarise the most important characteristics for a transfer protocol as:

\begin{enumerate} 
    \item The state transfer should be as close to perfect as possible (fidelity 1 between the states at the ends of the chain, $\ket{\psi(t=0)}_1$ and $\ket{\psi(t=\tau)}_N$).
    \item The state transfer should work at arbitrary distances.
    \item Preferably, the Hamiltonian should be as simple as possible and easily implementable in experiments.
\end{enumerate}

The first efforts to design state transfer protocols for quantum computing were done by Bose et al. \cite{bose_quantum_2003, burgarth_quantum_2007}. The simplest version consists of a one-dimensional spin chain with a Heisenberg Hamiltonian and constant uniform couplings between nearest-neighbour spins

\begin{equation}
    H = -J\sum_{<i,j>}\vec{\sigma^i}\cdot\vec{\sigma^j} -B\sum_{i=1}^{N}\sigma^i_z.
\end{equation}

This system can transfer a state from the first spin to the last one with fidelity 1 when $N\leq 4$, but for longer chains the fidelity decreases as $N^{-1/3}$. After that, several strategies were proposed aimed at keeping the fidelity at 1, all involving modifications to the couplings or the system's geometry. These strategies include using time-dependent couplings \cite{huang_quantum_2018}, multiple chains \cite{burgarth_conclusive_2005}, or "holes" at the ends of the chain \cite{gualdi_entanglement_2009, wojcik_unmodulated_2005}.

However, one strategy of particular interest for its accuracy and simplicity is to use constant but inhomogeneous couplings along the whole chain, as in Ref. \cite{christandl_perfect_2004}, which is the model on which we will base our protocol. In this case, the authors employ an XY Heisenberg Hamiltonian of the form

\begin{equation}
    H_G  =  -\sum_{n=1}^{N-1}\frac{t_n}{2}\left[\sigma_n^x\sigma_{n+1}^x + \sigma_n^y\sigma_{n+1}^y\right] \ ,
    \label{eq:hamiltonian_standard}
\end{equation}
where the couplings are chosen according to
\begin{equation}
    t_n = \frac{\lambda}{2}\sqrt{n(N-n)}.
    \label{eq:symmetry}
\end{equation}
 This makes the Hamiltonian identical to the angular momentum operator of a spin $S = \frac{1}{2}(N-1)$ particle, $H_G = \lambda S_x$. Note that the couplings in Eq. \eqref{eq:symmetry} show a mirror symmetry in the coupling strengths.

Under the Hamiltonian in Eq. \eqref{eq:hamiltonian_standard}, the probability amplitude of state transport between the two ends of the chain is periodic in time,

\begin{equation}
    \bra{00...01}e^{-iH_Gt}\ket{10...00} = \left[-i\sin\left(\frac{\lambda t}{2}\right)\right]^{N-1}.
    \label{eq:fidelity_standard}
\end{equation}

So perfect state transfer can be achieved at time $t = \tau = \pi/\lambda$ for any chain length. Note that this time is constant for any length of the chain $N$. This appears to be in contradiction with the intuition that any propagating effect must take longer to traverse a bigger system. However in this case the coupling strengths $t_n$ also scale with the system size $N$, meaning that the interactions are stronger for longer chains, and accelerate the transport velocity. Alternatively, one can decrease the parameter $\lambda$ linearly with $N$ and keep the coupling strengths below an acceptable bound, resulting in linearly growing transfer time. This is relevant for realistic systems, where the available coupling strengths are typically bounded.

\begin{figure*}
    \centering
    \includegraphics[width=2.0\columnwidth]{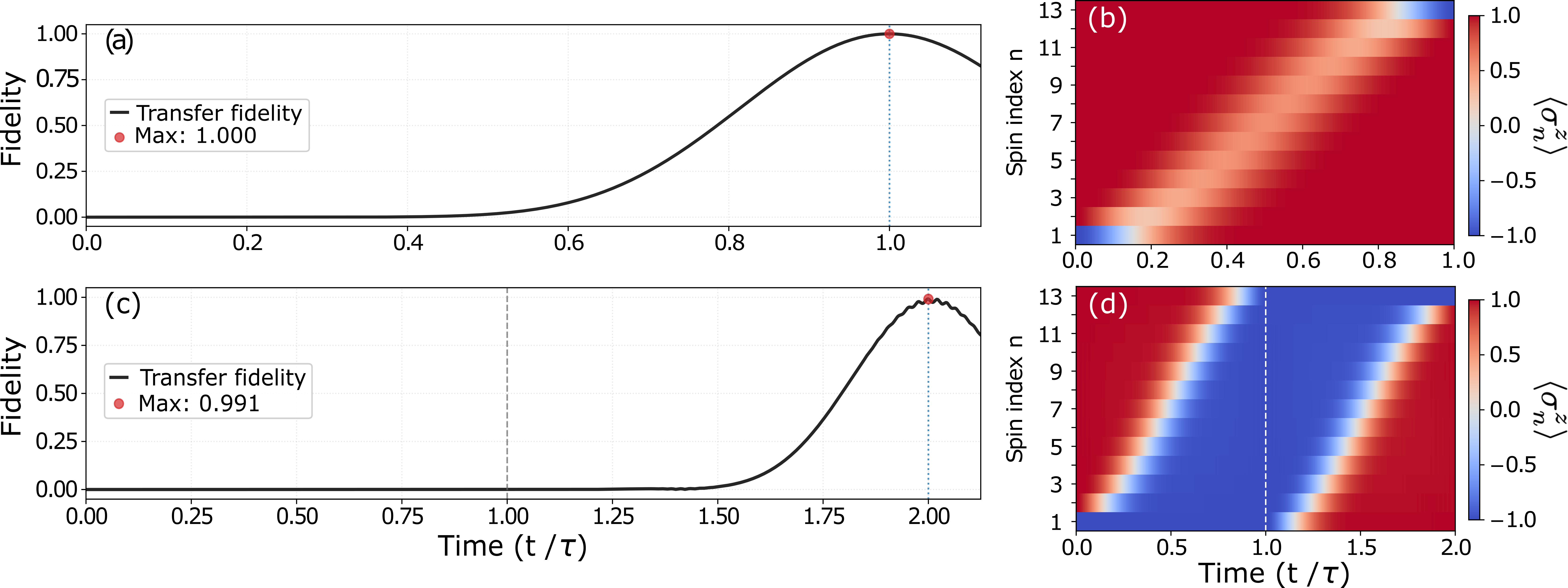}
    \caption{Figures (a, b) describe a spin transport using the standard spin-chain system Eq. \eqref{eq:hamiltonian_standard}. (a) Fidelity between the state at time $t$ and the final expected state $\ket{0...01}$, for $N=13$, and initial state $\ket{10...0}$. It reaches the maximum of 1 after a finite time $\tau$. (b) Evolution of z-component of each spin. The $-1$ value corresponds to the spin in the $\ket{1}$ state, and the $+1$ to the state $\ket{0}$. We can observe the swap of the initial and final spins after time $\tau$. Figures (c, d) describe the same transport, but using the domain wall encoding and the two-step protocol. (c) Fidelity between the state at time $t$ and the final expected state $\ket{0...01}$, for $N=13$, $J=0.5$ GHz, $\lambda = 22.72$ MHz, and initial state $\ket{10...0}$. (d) Evolution of z-component of each spin of the domain-wall system. We can observe the initial domain wall travelling to the other end of the chain after time $\tau$. However, there is now another domain wall traversing the chain from time $\tau$ to $2\tau$, due to the required reset of the chain.}
    \label{fig:standard_encoding_example}
\end{figure*}

Additionally, the Hamiltonian above commutes with the total angular momentum operator $Z = \sum_{n=1}^{N}\sigma^z_n$, meaning that it conserves the number of excitations, i.e. number of spins in state $\ket{1}$. This property will be relevant when we make modifications to this initial protocol. Figure \ref{fig:standard_encoding_example} (a, b) show an example of how the state $\ket{1}$ is transferred for a chain length of $N = 13$. We use two methods to visualise the results. One is by directly calculating the fidelity between the entire chain state at time $t$ and the expected final state, as can be seen in Figure \ref{fig:standard_encoding_example} (a). In the second one we plot the expectation value of the z-component $\langle \sigma^z \rangle$ for each spin in the chain and each time step, shown in Figure \ref{fig:standard_encoding_example} (b). This is a good visual way to see the state being transported, although for more complex states, this will not give as much information as the fidelity test. 

Even though the example from Eq. \eqref{eq:fidelity_standard} shows perfect state transfer, the property does not hold a priori for an arbitrary state of the first qubit, due to a relative phase of $(-i)^{N-1}$ between the $\ket{0}$ and $\ket{1}$ components of the final state  \cite{christandl_perfect_2005}. This phase difference can be corrected either after the transport by applying a rotation on the last qubit, or by adding local $Z$-terms to the Hamiltonian. 

The Hamiltonian from Eq. \eqref{eq:hamiltonian_standard} can also transfer superpositions of computational basis states with distinct excitation numbers $M$. However, there is an added subtlety, since components of the state with different $M$ will incur different local phases in the transport process. The local phases will depend on the $N$-dependent single-excitation phase mentioned above, as well as additional $M$-dependent phases due to exchange in the ordering of the excitations given by $(-1)^{M(M-1)/2}$. Since both phases are known, they can easily be corrected by either Alice or Bob.

\section{Domain wall encoding for simulating Heisenberg chains}
\label{sec:domain_walls}

The spin chain Eq. \eqref{eq:hamiltonian_standard} is the one we will use for our protocol. However, as of now it is not suited for implementation in Ising-like hardware, such as superconducting flux qubits. The issue is that it uses an XY Heisenberg Hamiltonian, and there is no clear way yet to recreate the XX and YY interactions with many analog devices, or the existing techniques have not yet been implemented in real hardware \cite{hita_perez_design_2025}. The way to circumvent this is to change the way in which we encode the logical spins into the physical system.

The standard encoding is to associate the state of each physical spin to the state of a logical spin. However, we can also encode the information using the so-called domain wall picture \cite{chancellor_domain_2019}. The main idea, drawn from classical magnetism, is to place the logical spin in the interface between two spins, and assign it the value $\ket{0}$ if those spins are in the same state, and $\ket{1}$, if they are in opposite states. As an example of this identification, observe the following states encoding one and two excitations respectively:

\begin{equation}
    \begin{aligned}
        \ket{111000}_{DW} \equiv \ket{00100}_{\text{standard}} \ , \\
        \ket{111011}_{DW} \equiv \ket{00110}_{\text{standard}} \ .
    \end{aligned}
    \label{eq:DWEncoding}
\end{equation}

On the right side of the identity, the '1' represents the transition between two domains of zeros and ones. Hence, the name domain wall encoding.

This encoding is applicable to the case of linear spin chains, and has already shown the ability to replicate several systems that were a priori not possible in many superconducting chips \cite{werner_quantum_2024}. The key advantage of this encoding is that, since we are changing the dynamics in the physical system, we also need to change the Hamiltonian. For a large class of Heisenberg Hamiltonians in the standard picture, their domain wall equivalent only contains ZZ interaction terms. In our quantum transport case, we switch from performing the operation $\ket{100...0} \rightarrow \ket{00...01}$ to performing $\ket{100...0} \rightarrow \ket{11...10}$. The Hamiltonian that achieves this result is

\begin{equation}
    H_{DW} = +\sum_{n=1}^{N}t_n\sigma_n^x - J\sigma_1^z + J\sigma_{N}^z + \sum_{n=1}^{N-1}J\sigma_n^z\sigma_{n+1}^z.
    \label{eq:Hamiltonian_domain_wall_1}
\end{equation}

This expression comes from considering a chain of spins containing domain walls, and constrained by a strong ferromagnetic coupling J. The $-J\sigma_1^z$ and $J\sigma_{N}^z$ represent local fields, similar to coupling the ends of the chain to fixed, virtual spins \cite{chancellor_domain_2019}. The opposite signs guarantee that there is at least one domain wall in the ground space, corresponding to one excitation in the chain.

In the Ising model Eq. \eqref{eq:Hamiltonian_domain_wall_1}, large ZZ-terms prevent the creation or destruction of domain walls, conserving the number of excitations. If $|J|$ is large enough, the system evolves in the subspace with a constant excitation number $M$. This restricts the dynamics to domain walls moving left and right. These dynamics are added through the transverse field terms $\sum_n t_n \sigma_n^x$, where $t_n$ are also given by Eq. \eqref{eq:symmetry}, i.e. they contain the same mirror symmetry relevant to our problem. These introduce movement of the domain walls along the chain by inducing spin flips at the domain wall. The spin flips can also create or destroy domain walls, but this effect is suppressed when the domain wall coupling is large $(|J|\gg \lambda)$.

Effectively, this system is equivalent to an energy transfer Hamiltonian, where excitations in a one-dimensional lattice are hopping from one site to the adjacent ones. The transverse field effectively acts as a hopping term on the domain walls, i.e.
\begin{equation}
    \sum_n \sigma^x_n \equiv \sum_n \sigma_{n+1}^+\sigma_n^- + \sigma_{n+1}^- \sigma_n^+ \ .
\end{equation}
The strength of the coupling $J$, however, is extremely relevant, since the map between the Heisenberg model from Eq. \eqref{eq:hamiltonian_standard} to the domain-wall Hamiltonian Eq. \eqref{eq:Hamiltonian_domain_wall_1} is exact for $|J| \to \infty$, and a finite $|J|$ is only an approximation.

An additional point to consider is that due to the ZZ-coupling term in Eq. \eqref{eq:Hamiltonian_domain_wall_1}, states with $M$ domain walls will pick up an additional phase as a result of time evolution under Hamiltonian Eq. \eqref{eq:Hamiltonian_domain_wall_1}. The reason is that due to the ZZ-interaction term, the $M$ domain wall subspace has an energy offset
\begin{equation}
    E_M = J(N-2M) \ .
    \label{eq:EnergyOffset}
\end{equation}
If the initial state only has support on computational basis states with the same $M$, this will merely result in a global phase. On the other hand, if the initial state is a superposition of states with distinct $M$, then there will be local phases, which we need to consider by correcting the phases by either Alice or Bob. This can always be done simply by evolving e.g. the received state by the

\begin{equation}
    U_{\text{offset-correction}}= \exp \left( i \tau J \sum_{n=1}^{N-1} \sigma_n^z \sigma_{n+1}^z \right)
    \label{eq:OffsetCorrection}
\end{equation}

From this expression, we can also see that if we chose $J$ such that $J\tau_{transfer}$ is a multiple of $2\pi$ the offset phase will be zero right at the completion time of the transport, which is known for our model ($\tau_{transfer} = \pi/\lambda$).

At the end of Section \ref{sec:standard_transfer}, we briefly mentioned how in the standard picture, arbitrary multi-qubit states can be transported at the cost of incurring correctable relative phases in components of the state with different excitation number $M$. Note that the local phases incurred due to the energy offset Eq. \eqref{eq:EnergyOffset} in the domain wall encoding are an additional independent phenomenon. Since it is also known, it can also be corrected.

The protocol by Christandl et al. \cite{christandl_perfect_2004} was originally intended for perfect state transfer. However, as we discussed above, the choice of domain wall encoding is only an approximation of the original protocol for finite $J$. This introduces errors in the transport process that we will now discuss.

Ref. \cite{werner_quantum_2024} employs the Schrieffer-Wolf transformation \cite{bravyi_schrieffer-wolff_2011} to estimate the fidelity of the approximate domain wall Hamiltonian. This method compares the exact Hamiltonian in the spin-chain picture with the domain wall one, and allows us to see their discrepancies as a function of the domain wall coupling $J$. A step-by-step derivation can be found in \cite{werner_quantum_2024}, and yields a leading order correction to the domain wall Hamiltonian of $O(|J|^{-1})$. Finally the fidelity between a state evolving under the effective domain wall Hamiltonian and the exact one is 

\begin{equation}
    \mathcal{F} = 1 - O(t^2|J|^{-2}).
\end{equation}

We see that for a fixed time $t=\tau$, such as the period of the state transfer protocol, the fidelity of the obtained state and the target state will be inversely proportional to the square of the domain wall coupling $J$.

On the other hand, by increasing $J$ we can reduce the error to an arbitrary quantity. However, the physical implementation of $J$ will be restricted by some upper bound, so at some point we will accumulate a substantial error. This however can be mitigated at the cost of a time overhead. We can see this by explicitly writing the parameter dependencies of Eq. \eqref{eq:Hamiltonian_domain_wall_1}, that is $H_{DW}(\vec{t}, J)$, where we have denoted $\vec{t}$ as the vector of transverse field strengths $t_n$ from Eq. \eqref{eq:symmetry}. Since all $t_n$ have a constant factor $\lambda$, we will take it out as $t_n = \lambda v_n$  Then, after some simulation time $t$, the Hamiltonian evolves as $e^{-itH_{DW}(\lambda\vec{v}, J)}$, and again we take the overall factor $\lambda$ and obtain $e^{-i\lambda tH_{DW}(\vec{v}, J/\lambda)}$. If $\lambda < 1$ with this we have essentially rescaled the domain wall coupling to a higher value at the cost of adding a time overhead. 

Then, the fidelity also depends on the parameter $\lambda$ such that

\begin{equation}
    \mathcal{F} = 1 -  O(t^2\lambda^{2}|J|^{-2}).
    \label{eq:infidelity}
\end{equation}

And thus, we can reduce the error by decreasing $\lambda$, so for a given error tolerance of $\epsilon = 1 - \mathcal{F}$, we can set $\lambda$ to be

\begin{equation}
    \lambda = O\left(\frac{|J|\sqrt{\epsilon}}{t}\right).
\end{equation}

The mathematical details of the mapping and the error analysis have been discussed in detail in Ref. \cite{werner_quantum_2024}.

\section{Two QST protocols with Ising Hamiltonians}
\label{sec:DW_protocol}

This section describes the step-by-step implementation of the state transfer protocol using domain walls. We will start from the simpler case of sending single qubits and then extend it to more complex states. All numerical results have been obtained using the QuTip library in Python \cite{lambert2024qutip5quantumtoolbox}. 

\subsection{Single-qubit transport}

Let $\ket{\psi}_1$ be an arbitrary state of Alice's spin in the chain, $\ket{\psi}_1 = \alpha\ket{1} + \beta\ket{0}$. The rest of the spins are in the state $\ket{0}$. Then the state of the whole chain is 

\begin{equation}
    \ket{\psi(0)} = \alpha\ket{100...00} + \beta\ket{000...00}.
    \label{eq:initial_state_standard}
\end{equation}

As mentioned above, the objective of the transport is to move the state of the first spin to the last one, so that the final state after time $\tau$ is

\begin{equation}
    \ket{\psi(\tau)} = \alpha\ket{000...01} + \beta\ket{000...00}.
    \label{eq:final_state_standard}
\end{equation}

In the domain wall picture, the evolution of the system will go from

\begin{equation}
    \ket{\psi(0)} = \alpha\ket{100...00} + \beta\ket{000...00}
    \label{eq:initial_state_dw}
\end{equation}

to

\begin{equation}
    \ket{\psi(\tau)} = e^{i\tau\phi_{\alpha}}\alpha\ket{11...10} + e^{i\tau\phi_{\beta}}\beta\ket{00...00}.
    \label{eq:middle_state_dw_1}
\end{equation}

In the non-zero state component of the final state, all spins except for the last one are in the state $\ket{1}$, meaning that the state of the logical spins in their interface between spins $N-1$ and $N$ is $\ket{1}$. Effectively, the initial excitation has moved to the end of the chain.

There is also the presence of additional phases proportional to $J$ coming from an energy offset $E_M$ between the effective Heisenberg Hamiltonian and the domain wall Hamiltonian given by Eq. \eqref{eq:EnergyOffset}.

This means that for the superposition Eq. \eqref{eq:middle_state_dw_1}, where one component has $M=0$ and the other $M=1$, there is a global phase $\phi_{glob} = JN\tau$, as well as a relative phase $\phi_{rel} = -2J\tau$. This relative phase will have to be taken into account when making measurements.  However, since it is a known quantity, it can be corrected either before sending the spins, or afterwards by the receiver using the unitary Eq. \eqref{eq:OffsetCorrection}.\\
We will rewrite Eq. \eqref{eq:middle_state_dw_1} as
\begin{equation}
        \ket{\psi(\tau)} = e^{iJN\tau}\left(e^{-i2J\tau}\alpha\ket{11...10} + \beta\ket{00...00}\right)
    \label{eq:middle_state_dw_2}
\end{equation}

\begin{figure}
    \centering
    \includegraphics[width=0.4\textwidth]{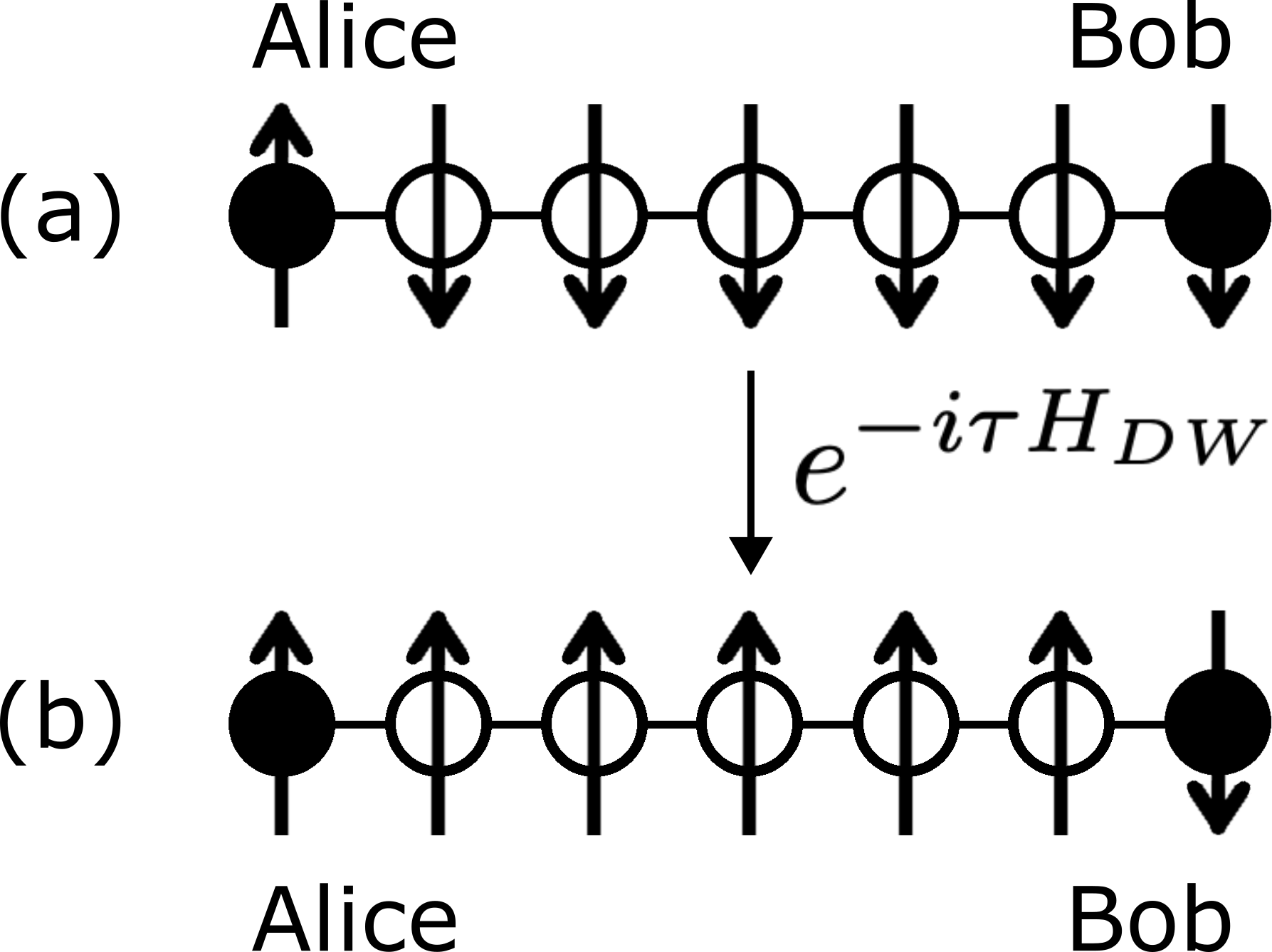} 
    \caption{Domain wall representation of the initial state $\ket{100000}$ (a) and final state $\ket{000001}$ (b) evolving under the Hamiltonian from Eq. \eqref{eq:Hamiltonian_domain_wall_1}. Initially, the domain wall is located between the first physical spin and the virtual, fixed spin. After the evolution, it has moved to the last physical and the virtual fixed spin at the other end of the chain.}
    \label{fig:diagram_classical}
\end{figure}

The Hamiltonian from Eq. \eqref{eq:Hamiltonian_domain_wall_1} is the naive translation of the state transfer problem Hamiltonian Eq. \eqref{eq:hamiltonian_standard}. However, note that the superposition of states in Eq. \eqref{eq:initial_state_dw} and Eq. \eqref{eq:middle_state_dw_2} are not in principle attainable by the Hamiltonian with Eq. \eqref{eq:Hamiltonian_domain_wall_1}. The reason is the strong ferromagnetic coupling to virtual spins at the ends of the chain. We can see a visual representation of these terms in Figure \ref{fig:diagram_classical}, which shows the initial and final states of the chain, and where the black spins represent the virtual spins at the extremes, i.e. the $\pm J$ local fields.

Since the virtual spins are realised by local z-fields, they are effectively classical spins. This means that Eq. \eqref{eq:Hamiltonian_domain_wall_1} only allows us to construct the state $\ket{1}$ at the start of the chain (in other words, $\alpha = 1, \beta = 0$), or, if the virtual spin is flipped, the state $\ket{0}$. However, we need to be able to construct a superposition of both states in the first domain wall. Otherwise we are effectively transporting classical information.

We solve the restriction by removing the virtual qubit in the first spin and encoding the information into the first physical spin instead. Also, to keep a ferromagnetic-like boundary condition we do not apply the transverse field to the first spin, effectively preventing it from evolving. In other words, we replace the fixed virtual spin with a fixed physical spin. The $\sigma^x$ Hamiltonian term starts from $n=2$ instead of $n=1$:

\begin{equation}
    H_{\text{transport}} = +\sum_{n=2}^{N}t_{n-1}\sigma_n^x + J\sigma_{N}^z + \sum_{n=1}^{N-1}J\sigma_n^z\sigma_{n+1}^z
    \label{eq:Hamiltonian_domain_wall_2}
\end{equation}

The fixing of the physical spin could potentially also be achieved by applying a strong local field, although in this case the system will pick up an additional known relative phase which needs to be considered by either sender or receiver. 

\begin{figure*}
    \centering
    \includegraphics[width=1.0\textwidth]{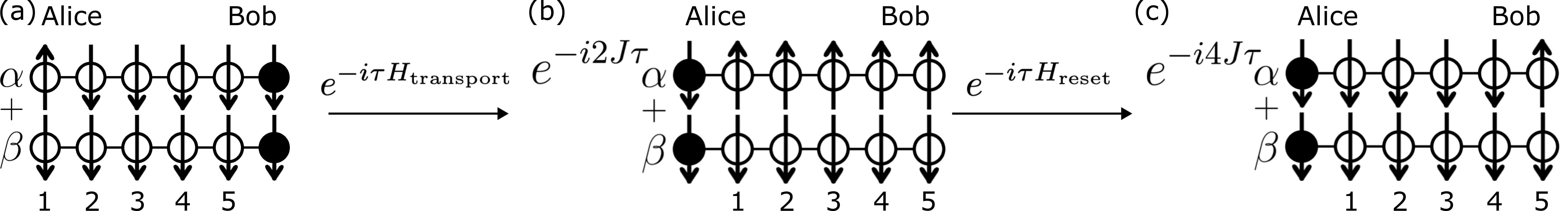} 
    \caption{(a) Domain wall representation of the state from Eq. \eqref{eq:Hamiltonian_domain_wall_2} under the domain wall Hamiltonian with only one virtual spin and no transverse field in the first physical spin; (b) Change of Hamiltonian after the transport has been completed, where the virtual spin is switched to the first spin and the transverse field is removed from the last spin. The relative phase between the states is also included; (c) End result of the transport after the Hamiltonian switch, including the relative phase between the states $\ket{1}$ and $\ket{0}$. The physical spins are numbered 1-5, while the fixed virtual spins are colored black.}
    \label{fig:initial_state_superposition}
\end{figure*}

With this, we can transport one qubit in an arbitrary state by preparing the first physical spin in the state $\ket{\psi} = \alpha \ket{1} + \beta \ket{0}$ to be transported. This new initial state is represented visually in Figure \ref{fig:initial_state_superposition} (a).

A second issue arises when looking at the spins along the chain. If we compare Eq. \eqref{eq:final_state_standard} and Eq. \eqref{eq:middle_state_dw_2}, we will observe that in the final state in the domain wall picture Eq. \eqref{eq:middle_state_dw_2}, all spins along the chain are potentially entangled, depending on the initial states $\alpha$ and $\beta$. That is, by measuring any spin along the chain, we are able to infer the state of the rest. This does not happen in the standard picture where only the two extremes contain information about the system.

This makes it so that the quantum information is delocalised across the whole chain, and we cannot extract the information about the phase of the state by only measuring the last spin. However, this can be solved by adding an additional step in the protocol to disentangle the chain. Essentially, we want it to evolve so that all the spins between the extremes are reset to the state $\ket{0}$. 

We achieve this effect by modifying the Hamiltonian after the state Eq. \eqref{eq:middle_state_dw_2} is reached, inserting a virtual spin in the 'down' state at the start of the chain (adding a $+J\sigma_z^1$ term through a local field), while at the same time removing the virtual spin that we had placed initially at the end of the chain. At the same time, we also remove the transverse field in the last physical spin. Once again, this is done to prevent it from evolving and storing the information that we had previously transported. Then we let the system evolve under this changed Hamiltonian defined as

\begin{equation}
    H_{\text{reset}} = +\sum_{n=1}^{N-1}t_n\sigma_n^x - J\sigma_1^z + \sum_{n=1}^{N-1}J\sigma_n^z\sigma_{n+1}^z \ .
    \label{eq:Hamiltonian_domain_wall_3}
\end{equation}

The end result of this operation will be that the whole chain except the last spin will "flip down" leaving us with the target state in the last spin,

\begin{equation}
    \ket{\psi(2\tau)} = e^{i2JN\tau}  \left(e^{-i4J\tau}\alpha\ket{000...01} + \beta\ket{000...00}\right),
\end{equation}
 
which is the same as Eq. \eqref{eq:final_state_standard} save for the global and relative phases. The changes in Hamiltonian and the final state can also be seen in Figures \ref{fig:initial_state_superposition} (a-c), including the relative phases.

The main cost of this operation is that we double the transfer time, since it now involves two steps. Additionally, we are forced to change the Hamiltonian and displace the transverse fields in the spins at a precise time (in order to preserve the mirror symmetry), which could pose a hardware challenge when implementing it.

An analog example to Figures \ref{fig:standard_encoding_example} (a, b) showing the result of transporting the same state can be seen in Figures \ref{fig:standard_encoding_example} (c, d). The fidelity plot Figure \ref{fig:standard_encoding_example} (c) is very similar to Figure \ref{fig:standard_encoding_example} (a), except for the extended time axis and the small oscillations near the peak. This is due to couplings to states outside the subspaces of $M=0,1$, and their effect is suppressed as $J$ approaches infinity, as per Eq. \eqref{eq:infidelity}. This is also the reason for the fidelity to reach close to, but not exactly, unity. We will also verify the reduction of the error below. Additionally, Figure \ref{fig:standard_encoding_example} (d) highlights the difference in the chain dynamics with the Heisenberg case. Note that the switch of the Hamiltonian from $H_{\text{transport}}$ to $H_{\text{reset}}$ happens at $t/\tau = 1$, denoted by a dashed line in Figure \ref{fig:standard_encoding_example} (c).

\subsection{Multi-qubit transport}

With the transport protocol established, we can extend it to states with more than one qubit. In this case we will divide the whole chain in three sections. The first and second are the registers of Alice and Bob, which will contain $N$ physical spins (to transport an $N$-qubit state). Finally, there is the wire section of the chain, which in principle can have an arbitrary length. The first phase of the transport will work in the same way, by adding a virtual spin to the end of the chain and removing the transverse $\sigma_x$-field in the first spin. Then the state in Alice's register will evolve to form a mirror image in Bob's register. At that point, we do several things. First, we disable the transverse field in Bob's register. We do so to prevent those spins from evolving. This way, the transferred state can be measured from Bob's register once the rest of the chain has been set to $\ket{0}$. Next, we reconfigure the transverse fields in Alice's register and the wire, so that they fulfill the mirror symmetry Eq. \eqref{eq:symmetry}, i.e. the reset Hamiltonian $H_{\text{reset}}$ only acts on Alice's register and the wire
\begin{equation}
    H_{\text{reset}} = \sum_{n=1}^{N_\text{Alice} + N_\text{wire}} t_n \sigma_n^x  -J \sigma_1^z +  \sum_{n=1}^{N-1}J\sigma_n^z\sigma_{n+1}^z \ ,
    \label{eq:Hamiltonian_domain_wall_4}
\end{equation}
where $N_\text{Alice}$ and $N_\text{wire}$ are the number of spins on Alice's register and the wire respectively.

\begin{figure}
    \centering
    \includegraphics[width=\columnwidth]{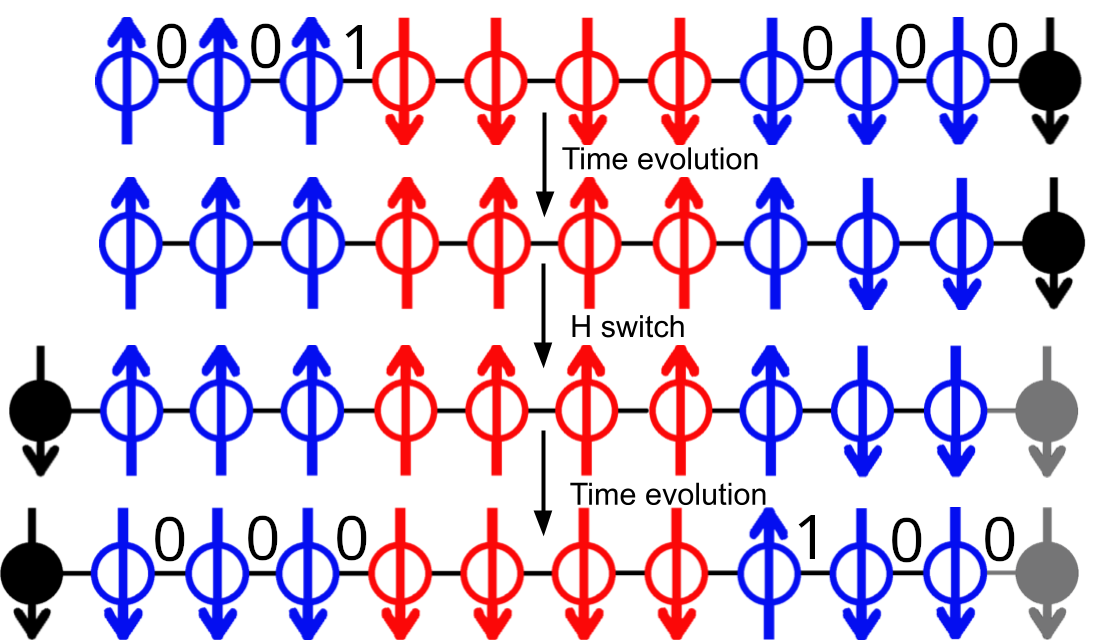} 
    \caption{Example of an initial chain for the transfer of the logical state $\ket{001}$. The blue sections left and right are Alice and Bob's registers respectively, and the red section is the length of the wire. The different stages of the protocol are shown with the changes in the virtual spins and the results of time evolution. The gray spin at the right represents a field that is turned off, but is still used to determine the value of the last logical spin we measure ($\ket{0}$ if the last physical spin is down and $\ket{1}$ if it is up).}
    \label{fig:multiqubit}
\end{figure}

Finally, we "move" the virtual spin to the start of the chain, resetting all spins to the "down" state. The difference with respect to the one-qubit case is that in the resetting step the entirety of Bob's register is not affected by the transverse field, as opposed to only the last qubit, and also we rearrange the transverse field terms along the whole chain, as opposed to just shifting them by one qubit. Note that these operations correspond to turning off and on local fields at a predetermined time and does not require sending information between Alice and Bob. Figure \ref{fig:multiqubit} shows a diagrammatical representation of these steps.

There is one detail to be considered by Alice when encoding the quantum data into a domain-wall state. As discussed in Ref. \cite{werner_quantum_2024}, the domain wall encoding has a $\mathbb{Z}_2$-symmetry, since $\ket{\uparrow \downarrow}$ and $\ket{\downarrow \uparrow}$ ($\ket{\downarrow \downarrow}$ and $\ket{\uparrow \uparrow}$) encode the same logical state. This leads in principle to an ambiguity with regard to the parity, however, since Alice's register is in contact with the wire and the wire is assumed to be in the all-down state initially, they need to chose the encoding of their state that respects this constraint. As an example, consider the logical state $\ket{10}$, which in the domain-wall encoding could be encoded as both $\ket{\uparrow \downarrow \downarrow}$ and $\ket{\downarrow \uparrow \uparrow}$. However, if we consider the wire as well, we see that
\begin{equation}
    | \underbrace{\uparrow \downarrow \downarrow}_{\text{register}} \underbrace{\downarrow \downarrow ...}_{\text{wire}} \rangle \not\equiv | \underbrace{\downarrow \uparrow \uparrow}_{\text{register}} \underbrace{\downarrow \downarrow ...}_{\text{wire}} \rangle \ ,
    \label{eq:ParityConstraint}
\end{equation}
since at the boundary of register and wire on the right-hand side of Eq. \eqref{eq:ParityConstraint}, there is another domain wall introduced. However, for all logical computational basis states, one of the two possible encodings always respects the constraint and thus it can also be respected for superpositions.

\begin{figure}
    \centering
    \includegraphics[width=\columnwidth]{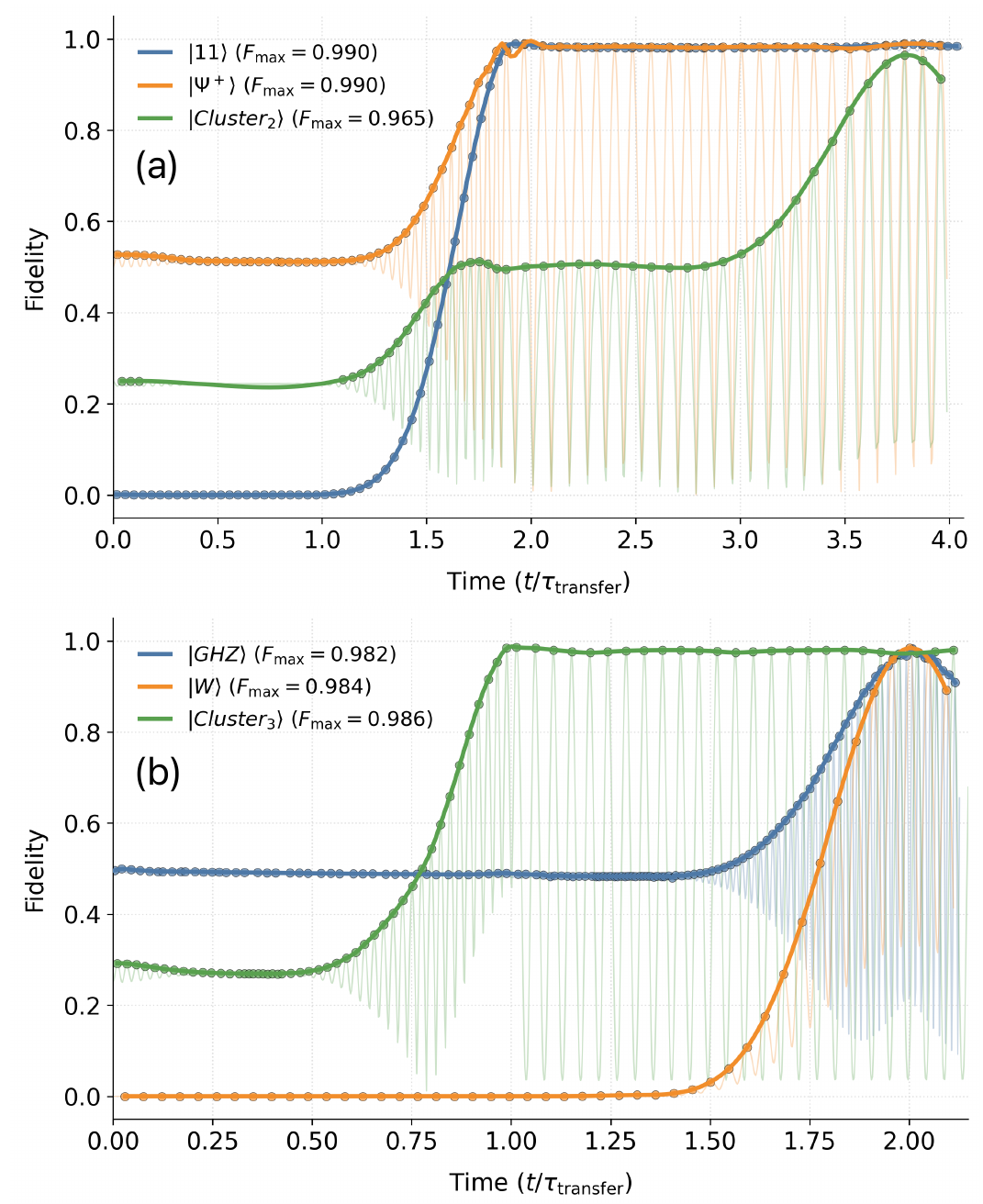} 
    \caption{Fidelity for several 2-qubit states (a) and 3-qubit states (b). Fidelities have fast oscillations given by the relative phase that the energy offset from Eq. \eqref{eq:EnergyOffset} introduces in states with different numbers of excitations. The peaks are marked by the dots, with interpolation lines linking them. On the upper plot we show the pure state $\ket{11}$ (blue), the Bell state $\ket{\psi^+} = \frac{1}{\sqrt{2}}(\ket{11} + \ket{00})$ (orange), and the cluster state $\ket{C_2} = \frac{1}{2}(\ket{00}+\ket{01}+\ket{10}-\ket{11})$ (green). At the bottom plot we show the following 3-qubit examples: A GHZ state, $\ket{GHZ} = \frac{1}{\sqrt{2}}(\ket{111} + \ket{000})$ (blue), the $W$ state $W = \frac{1}{\sqrt{3}}(\ket{001} + \ket{010} + \ket{100})$ (orange), and another cluster state $\frac{1}{2}(\ket{000}+\ket{011}+\ket{101}-\ket{110})$ (green).}
    \label{fig:fidelities_2-3}
\end{figure}

Figure \ref{fig:fidelities_2-3} shows the fidelity for examples of 2- and 3-qubit states. We chose these states to show the capability of these chains to transfer states with and without entanglement, and with applications of interest. For example GHZ states are important for fault-tolerant quantum computing \cite{yimsiriwattana_generalized_2004}, $W$ states are relevant for quantum networks and interferometry \cite{jordan_optimization_2025}, and the cluster states can be used for measurement-based quantum computing \cite{raussendorf_measurement-based_2003}, or teleportation protocols \cite{zeng_quantum_2003}. 

An important observation about these results is the fast oscillations that make it hard to determine the point where the transfer is successful. This effect is attributed to the encoding scheme and the fact that we use a finite ZZ coupling between spins. First, the finite coupling accumulates errors the more domain walls we have in the chain, and second there are phase-driven oscillations from the energy offsets Eq.\eqref{eq:EnergyOffset} that the domain wall encoding introduces. These phase differences are linear with $J$ so they are more pronounced for large coupling strengths. This could potentially make the measurement of the spin phases tricky as the time window where they are correct is small, requiring precisely timed measurements. However, if we account for these phases in the post-processing of the data we can mitigate or even correct their effect. In Figure \ref{fig:fidelities_2-3}, the thick graphs indicate the fidelity with correction of the energy-offset phase oscillations, while the shaded graphs show the fidelity without any correction.

Another relevant observation of the fidelities in Figure \ref{fig:fidelities_2-3} is that for some states, it reaches unity already after the transfer time $\tau$ and the reset step is superfluous. This is a result of the domain-wall encoding, as some states already have the correct parity, such that after the mirror image of the state reaches the other end of the chain, the wire is already left in the all-down state. In this case, the reset step does not do anything, other than adding phases. However, generally, it is unknown whether the reset step is acting trivially or not.

\subsection{One-step protocol}
\label{sec:one_step_version}

In the previous sections, we have motivated and shown the need to build our protocol as a two-step process: First, we send a state along the chain, and then we reset the chain by switching the Hamiltonian, all while preserving the transported information in Bob's register. This leads to our protocol needing double the time to complete, and it makes the Hamiltonian time-dependent, due to this need to change it exactly half-way through the process.

However, we have seen that for some states (\textit{e.g.} the Bell state of Figure \ref{fig:fidelities_2-3}) the reset step is superfluous when transporting an even number of domain walls, signaling that for some states, only the first phase of the protocol is needed, and the Hamiltonian can be time-independent. One can ask then if we could achieve the same effect for any arbitrary state. The solution is simple, to engineer the state preparation in such a way that there is always an even number of domain walls. We can do this by designating one qubit of Alice's register as an auxiliary spin which state always forces an even $M$. For example in the Bell state of Figure \ref{fig:fidelities_2-3} it would take the same state as the last spin, which would equate to an additional "0" in the state (and thus, no extra domain wall), but for the GHZ state of the same figure, it would be prepared in such a way to add a final domain wall, making the number always even:

\begin{equation}
\begin{split}
    \ket{\psi}_{Alice} = \ket{\Psi^+}  = &\frac{1}{\sqrt{2}}\left(\ket{11} + \ket{00}\right) \\ \xrightarrow[]{one-step} &\frac{1}{\sqrt{2}}\left(\ket{110} + \ket{000}\right) \\
    \ket{\psi}_{Alice} = \ket{GHZ}= &\frac{1}{\sqrt{2}}\left(\ket{111} + \ket{000}\right) \\ \xrightarrow[]{one-step} &\frac{1}{\sqrt{2}}\left(\ket{1111} + \ket{0000}\right)
\end{split}
\end{equation}

Finally, note that after the state transfer process, this extra qubit will also be transported to Bob's end, and thus has to be taken into account. However, since it is not part of the original state we intended to transfer, we have to ignore it in such a way that we recover all the information about our state. By that we mean that we cannot simply trace it out, since it contains phase information. This can be easily seen with an example: Assume we wish to transport the state $ | \psi \rangle$. The state can be decomposed as

\begin{equation}
    \ket{\psi} = \ket{\psi_0} + \ket{\psi_1} \ ,
\end{equation}
where $|\psi_0\rangle$ ($|\psi_1\rangle$) are supported on the computational basis states with an even (odd) number of domain walls. Note that the components $|\psi_0 \rangle$ and $|\psi_1 \rangle$ are orthogonal, but not necessarily normalised.

With the additional logical parity qubit, we end up transmitting the state $|\tilde{\psi} \rangle$, which can be decomposed into two components depending on the state of the parity qubit

\begin{equation}
    |\tilde{\psi} \rangle = \ket{\psi_0}\otimes\ket{0} + \ket{\psi_1}\otimes\ket{1} \ .
    \label{eq:decomposition_plus}
\end{equation}

Assume Bob would like to measure the expectation value of an observable $\hat{O}$ on the original state, i.e.

\begin{equation}
\begin{split}
    \bra{\psi} \hat{O} \ket{\psi} = & \bra{\psi_0} \hat{O} \ket{\psi_0} + \bra{\psi_1} \hat{O} \ket{\psi_1} \\
    & + 2 \Re \left( \bra{\psi_0} \hat{O} \ket{\psi_1} \right) \ .
    \label{eq:TrueExpectation}
\end{split}
\end{equation}

At the same time, if we were to ignore the parity qubit in the transmitted state $| \tilde{\psi} \rangle$, we would find that

\begin{equation}
    \begin{split}
        &\langle \tilde{\psi} | \mathbb{I} \otimes \hat{O} | \tilde{\psi} \rangle \\
        &= \langle 0 | 0 \rangle \bra{\psi_0} \hat{O} \ket{\psi_0} + \langle 1 | 1 \rangle \bra{\psi_1} \hat{O} \ket{\psi_1} \\
    & + 2 \Re \left( \langle 0 | 1 \rangle \bra{\psi_0} \hat{O} \ket{\psi_1} \right) \\
    &= \bra{\psi_0} \hat{O} \ket{\psi_0} + \bra{\psi_1} \hat{O} \ket{\psi_1}
    \end{split}
    \label{eq:NaiveExpectation}
\end{equation}

due to the orthonormality of $\ket{0}$ and $\ket{1}$. Generally, the expectation values in Eq. \eqref{eq:TrueExpectation} and Eq. \eqref{eq:NaiveExpectation} are not the same due to the missing contribution of the off-diagonal part of the observable.

This problem can easily be fixed by mapping the observable $\hat{O}$ to $\hat{O}' := (\mathbb{I} + \sigma^x) \otimes \hat{O}$. Simple algebra will show that

\begin{equation}
    \begin{split}
        \langle \tilde{\psi} | \hat{O}' | \tilde{\psi} \rangle = \bra{\psi} \hat{O} \ket{\psi} \ .
    \end{split}
\end{equation}

One subtlety about this scheme is how to actually implement this measurement, since this $\hat{O}'$ operator in the logical qubit space needs to be properly translated into the domain wall picture. Appendix A and B of Reference \cite{werner_quantum_2024} contain a detailed explanation of this mapping, however, it assumes that observables need to be expressed in Pauli strings of even length, which does not apply for the $(I + \sigma^x)$ part of our operator. This, however, is not a limiting factor in our case, the reason being that we relax the fixed spin conditions at the ends of the chain. A fully detailed reasoning can be found in Appendix \ref{sec:Appendix}.

\subsection{Error analysis}

\label{sec:errors}
In the previous simulations, we could already observe some error in the state transfer. Note that this error is a result of the approximate domain-wall encoding. The original QST protocol as designed for Heisenberg chains is in principle capable of perfect transfer. Hence, we would expect that the error of the transfer can be reduced by employing stronger couplings $J$, or, alternatively, re-scale the model parameters and extend the transfer time, as has been discussed above.

\begin{figure}
    \centering
    \includegraphics[width=1\columnwidth]{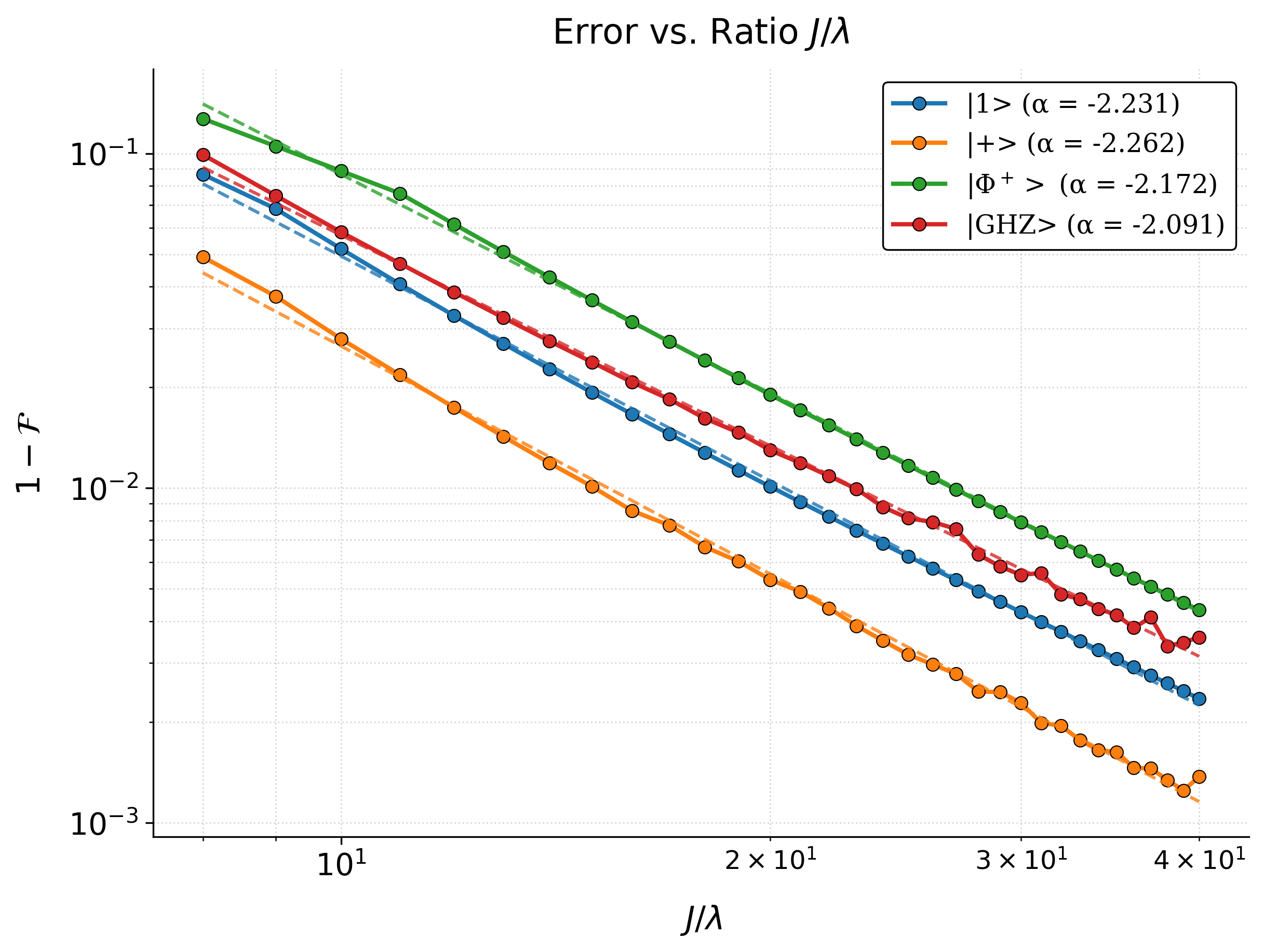} 
    \caption{Error for the transport as a function of $J/\lambda$ in a logarithmic scale for different example states. $\alpha$ represents the slope of the linear fit, which confirms the quadratic dependence of Eq. \eqref{eq:infidelity} on the model's parameters. The points contain values of $J/\lambda$ between 8 and 40. For lower numbers of the ratio, the higher order corrections to the domain wall error become large and the quadratic relation is broken.}
    \label{fig:linear_dependance}
\end{figure}

To analyze the error scaling we measure the fidelity of the QST in the domain wall picture as a function of the ratio $J/\lambda$. The plot of Figure \ref{fig:linear_dependance} shows the dependence of the infidelity $1-\mathcal{F}$ as a function of $|J|/\lambda$ for different states. As predicted by the theoretical analysis, the error decays with the square of the ratio $|J|/\lambda$. This gives us two parameters we can use to improve the quality of the QST up to a required precision. Thus, while our protocol is an imperfect implementation of a perfect QST protocol, the error can be arbitrarily reduced.

Note that here we analyze the error scaling of transferring individual states, which matches the theoretical bound Eq. \eqref{eq:infidelity}. Since the bound considers the worst case, it will also hold in the average case \cite{apollaro_quantum_2022}.

\subsection{Summary of the quantum transport protocol}
Here we describe our proposed protocol step-by-step. We consider a scenario where Alice wants to send a (multi-)qubit state $\ket{\psi}$ to Bob via a chain of spins with Ising interactions. This description matches the illustration in Figure \ref{fig:multiqubit}.

QST protocol with Ising Hamiltonians:
\begin{enumerate}
    \item The chain spins, as well as Bob's register are all in the "down" state, Bob couples their last physical spin to a virtual "down"-spin (i.e. turns on a local field)
    \item Alice prepares the domain wall representation of $\ket{\psi}$ in their register, respecting the parity constraint.
    \item The whole system evolves according to Hamiltonian $H_{\text{transport}}$ Eq. \eqref{eq:Hamiltonian_domain_wall_2} for time $\tau = \frac{\pi}{\lambda}$.
    \item Bob turns off the transverse field on their register, Alice couples a virtual "down"-spin to the end of their register, effectively introducing a new domain wall
    \item Alice's register and the chain evolve under the Hamiltonian $H_{\text{reset}}$ Eq. \eqref{eq:Hamiltonian_domain_wall_4} for time $\tau$.
    \item Bob applies $U_{\text{offset-correction}}$ Eq. \eqref{eq:OffsetCorrection} to his register and decodes the domain-wall state (or further processes the quantum data).
\end{enumerate}

Alternatively, if we use the one-step protocol using an additional qubit to make the excitation number even, the process is the same omitting steps 4 and 5.

Note that the encoding and decoding of an arbitrary logical state to and from the domain wall picture might require global operations on the Alice's and Bob's register, respectively. However, this is in accordance with the rules of QST, as we discussed in section \ref{sec:standard_transfer}, where we assume that Alice and Bob have full control of their registers and have no, or limited, control of the quantum wire between them.

Note further, that the changes to the Hamiltonian are independent of the quantum data that is sent and occur at predefined time, thus this step does not violate the QST rules either.
\section{Conclusion}
\label{sec:conclusions}

We have proposed a working quantum state transfer protocol using Ising Hamiltonians, based on a protocol by Christandl et al. \cite{christandl_perfect_2004}, which demonstrates perfect state transfer using a Heisenberg Hamiltonian with inhomogeneous XX+YY couplings. Our protocol avoids the use of XX and YY interaction terms, which are not feasible for many current analog quantum simulation platforms, by encoding the information in domain walls inspired by the results from Ref. \cite{werner_quantum_2024}. The result is a mapping of the exact Hamiltonian to an effective one.

One particularity of the protocol is that it contains an extra step that resets the chain to the state $\ket{0}$ after the transfer is complete by switching the virtual spins and redistributing the transverse fields, which disentangles the whole chain. However, by realizing that entanglement does not occur for states of even parity, we can avoid the extra step by using an auxiliary qubit that forces the excitation number to be even. Additionally, superpositions of states with different excitation numbers carry a known relative phase throughout the time evolution, which can be corrected.

Thus, we designed the protocol to be suitable for the transport of arbitrary superpositions of one-qubit states and multi-qubit states, and we have shown this numerically by applying it to relevant examples of up to 3-qubit states. We have validated the results by analyzing 1) the evolution of the Z-component of each spin and 2) the fidelity between the simulated chain and the theoretical expected results.

We have analyzed the error induced by the finite domain wall coupling, which has a dependence of $|J|^{-2}$ on the fidelity. Since the construction of the Hamiltonian allows for a rescaling of the parameters, we have shown that we can mitigate the error by effectively reducing the transverse field strength while adding a time overhead.

Although we have proposed a first working protocol that transfers quantum information with high accuracy, there are several potential next steps. First, several other protocols for perfect state transfer have been proposed \cite{burgarth_conclusive_2005, gualdi_entanglement_2009, huang_quantum_2018} which use variations of the Heisenberg Hamiltonian, albeit with additional features like time-dependent couplings, uncoupled spins, or different chain structures. It could be of interest to rewrite such protocols in the domain wall picture and study their speed, accuracy, and error tolerance to compare them against our proposal. Second, while we chose to incorporate domain walls as our method for obtaining a Hamiltonian with only ZZ interactions, as we have shown this approximation induces errors. We could consider alternative solutions and compare their impact on the total protocol error.

Finally, it would be extremely interesting to test the protocol on analog quantum hardware such as superconducting flux qubits with ZZ-couplings and to analyze the actual level of control needed to implement it efficiently. This would be a natural continuation of our work. However, there are some challenges, including the implementation of large coupling strengths, and the precise mirror symmetric transverse fields in all qubits. In devices where these shortcomings are addressed, our protocol could provide a practical mechanism for transporting quantum information within QPUs with short transfer times and high fidelity. Another potential use is a type of quantum repeater. This is possible because our protocol does not require the preparation of any state outside the initial one and includes a reset mechanism, allowing the system to periodically receive states from a sender and transfer them along the chain.
\section{Acknowledgements}
The authors thank Marta P. Estarellas and Artur García-Sáez for helpful discussions. MW acknowledges support by the European Commission EIC-Transition project RoCCQeT (Grant No. GA 101112839) and the Agencia de Gestió d'Ajuts Universitaris i de Recerca through the DI grant (Grant No. DI75). The authors thankfully acknowledge RES resources provided by Barcelona Supercomputing Center in MareNostrum 5 to FI-2025-1-0043.\\
OM and MW contributed equally to this work.

\appendix
\section{Mapping operators to the domain wall picture}
\label{sec:Appendix}
Consider a spin chain with $N$ sites with indices ranging from $1$ to $N$. The system is mapped to an Ising chain with $N+1$ spins with indices running from $1$ to $N+1$, however, depending on the context, spins $1$ or $N+1$ may be virtual spins.\\
To map observables to the domain wall picture, note that any observable $\hat{O}$ can be decomposed into Pauli-strings as
\begin{equation}
    \hat{O} = \sum_{\hat{P} \in \{\mathbb{I}, \sigma^x, \sigma^y, \sigma^z\}^{\otimes N} } a_P \hat{P} \ .
\end{equation}
In the domain wall picture, a logical $\sigma^z_i$ is mapped to a two-body operator on the Ising spins
\begin{equation}
    \sigma^z_i \equiv \sigma^z_i \sigma^z_{i+1} \ .
    \label{eq:DWZ}
\end{equation}
Additionally, we can map $\sigma^x_i$ to the domain wall picture
\begin{equation}
    \sigma^x_{i} \equiv \prod_{n = 1}^{i} \sigma_n^x
    \label{eq:DWX}
\end{equation}
acting on the physical spins. The right-hand side in Eq. \eqref{eq:DWZ} and Eq. \eqref{eq:DWX} are operators acting on the Ising spins. Furthermore, note that the Ising chain operator in Eq. \eqref{eq:DWX} may require flipping of the the virtual spin.\\
Using further that
\begin{equation}
     \sigma^y = i \sigma^x \sigma^z \equiv i \left( \prod_{n = 1}^{i} \sigma_n^x \right) \ \sigma^z_i \sigma^z_{i+1},
    \label{eq:DWY}
\end{equation}
we can construct all Pauli-operators in the domain wall picture and thus also map any observable by combinations of Eq. \eqref{eq:DWZ}, Eq. \eqref{eq:DWX} and Eq. \eqref{eq:DWY}.\\
Note that in Ref. \cite{werner_quantum_2024}, only the mapping of operators that decompose into strings of even length was considered. Here we overcome this limitation by also flipping the virtual spins, if there are any.

\bibliographystyle{unsrt}
\bibliography{references}

\end{document}